\documentclass[journal]{IEEEtran}
\makeatletter

\usepackage{blindtext}
\usepackage[T1]{fontenc}
\usepackage{textcomp}
\usepackage{amsmath}
\usepackage{amssymb}
\usepackage{bm}
\usepackage{mdwmath}
\usepackage{cases}
\usepackage[short, c2]{optidef}
\usepackage{graphicx}
\graphicspath{{Figure/PDF/}{Biography/PDF/}}
\usepackage[dvipsnames]{xcolor}
\definecolor{myblue}{rgb}{0,0.4980,1} 
\definecolor{myred}{rgb}{0.8706,0.1608,0.0627} 
\usepackage[caption=false,font=footnotesize,subrefformat=parens]{subfig}
\usepackage[nosort,noadjust]{cite}
\usepackage{url}
\usepackage{tabularray}
\UseTblrLibrary{counter}
\usepackage[linesnumbered,ruled]{algorithm2e}

\usepackage{hyperref}
\newcommand{\colorhypersetup}{\@ifpackageloaded{hyperref}{\hypersetup{%
	bookmarksopen=true,%
	bookmarksnumbered=true,%
	pdfpagemode={UseOutlines},
	pdfstartview={FitH},%
	colorlinks=true,%
	linkcolor={myred},%
	citecolor={orange}
}}{\empty}}
\newcommand{\blackhypersetup}{\@ifpackageloaded{hyperref}{\hypersetup{%
	bookmarksopen=true,%
	bookmarksnumbered=true,%
	pdfpagemode={UseOutlines},
	pdfstartview={FitH},%
	colorlinks=true,%
	allcolors={black}
}}{\empty}}
 \blackhypersetup

\usepackage{mfirstuc-english}
\MFUhyphentrue
\usepackage[version3,description,IEEEtran]{eacro}
\acsetup{%
    pages/display=all,%
    pages/seq/use=false,%
    pages/name=true,%
    pages/fill={\quad},%
    make-links=false%
}

\DeclareAcronym{bs}{
	short = BS,
	long = base station}
 \DeclareAcronym{mg}{
	short = MG,
	long = multicast group}
  \DeclareAcronym{dt}{
	short = DT,
	long = digital twin}
 \DeclareAcronym{lstm}{
	short = LSTM,
	long = longs short-term memory}
  \DeclareAcronym{rnn}{
	short = RNN,
	long = recurrent neural networks}
   \DeclareAcronym{ddqn}{
	short = DDQN,
	long = double deep Q-network}
 \DeclareAcronym{rmse}{
	short = RMSE,
	long = root mean square error}
 \DeclareAcronym{svc}{
	short = SVC,
	long = scalable video coding}
 \DeclareAcronym{msvs}{
	short = MSVS,
	long = multicast short video streaming}
\DeclareAcronym{td3}{
	short = TD3,
	long =  twin delayed deep deterministic policy gradient}
 \DeclareAcronym{mdp}{
	short = MDP,
	long =  Markov decision process}
 \DeclareAcronym{dt3}{
	short = DFTD3,
	long =  diffusion-based TD3}
\DeclareAcronym{smg}{
	short = SMG,
	long =  sub-MG}
 \DeclareAcronym{drl}{
	short = DRL,
	long =  deep reinforcement learning}

\newcommand{\upperroman}[1]{\uppercase\expandafter{\romannumeral#1}}

\newcommand{\myunit}[1]{%
	\ifmmode
		\mathrm{#1}
	\else
		$ \mathrm{#1} $
	\fi}
\newcommand{\murm}{%
	\ifmmode
		\text{\textmu}
	\else
		\textmu
	\fi}

\newcommand{\MYnewpage}{%
	\ifCLASSOPTIONonecolumn
		\ifCLASSOPTIONjournal
			\typeout{The onecolumn journal mode.}
			\newpage
		\fi
	\fi}

\newlength{\mysinglefigwidth}
\newlength{\mymultifigwidth}
\ifCLASSOPTIONonecolumn
	\AtBeginDocument{\setlength{\mysinglefigwidth}{0.7\linewidth}}
\else
	\AtBeginDocument{\setlength{\mysinglefigwidth}{\linewidth}}
\fi
\ifCLASSOPTIONonecolumn
	\AtBeginDocument{\setlength{\mymultifigwidth}{0.5\linewidth}}
\else
	\AtBeginDocument{\setlength{\mymultifigwidth}{\linewidth}}
\fi

\makeatother
\begin{document}
\title{Efficient Digital Twin Data Processing for Low-Latency Multicast Short Video Streaming}
\author{
	\IEEEauthorblockN{Xinyu Huang\IEEEauthorrefmark{1}, Shisheng Hu\IEEEauthorrefmark{1}, Mushu Li\IEEEauthorrefmark{2}, Cheng Huang\IEEEauthorrefmark{4}, and Xuemin (Sherman) Shen\IEEEauthorrefmark{1}}
	    \IEEEauthorblockA{\IEEEauthorrefmark{1}Department~of~Electrical~\&~Computer~Engineering,~University~of~Waterloo,~Canada\\
     \IEEEauthorrefmark{2} Department of Electrical, Computer, \& Biomedical
Engineering, Toronto Metropolitan University, Canada\\
     \IEEEauthorrefmark{4}School of Computer Science, Fudan University, China
     \\Email: \{x357huan, s97hu, sshen\}@uwaterloo.ca, mushu.li@ryerson.ca, chuang@fudan.edu.cn}

   }

\ifCLASSOPTIONonecolumn
    \typeout{The onecolumn mode.}

\else
    \typeout{The twocolumn mode.}
    
\fi

\ifCLASSOPTIONonecolumn
	\typeout{The onecolumn mode.}
\else
	\typeout{The twocolumn mode.}
\fi

\maketitle

\ifCLASSOPTIONonecolumn
	\typeout{The onecolumn mode.}
	\vspace*{-50pt}
\else
	\typeout{The twocolumn mode.}
\fi
\begin{abstract}
In this paper, we propose a novel efficient digital twin (DT) data processing scheme to reduce service latency for multicast short video streaming. Particularly, DT is constructed to emulate and analyze user status for multicast group update and swipe feature abstraction. Then, a precise measurement model of DT data processing is developed to characterize the relationship among DT model size, user dynamics, and user clustering accuracy. A service latency model, consisting of DT data processing delay, video transcoding delay, and multicast transmission delay, is constructed by incorporating the impact of user clustering accuracy. Finally, a joint optimization problem of DT model size selection and bandwidth allocation is formulated to minimize the service latency. To efficiently solve this problem, a diffusion-based resource management algorithm is proposed, which utilizes the denoising technique to improve the action-generation process in the deep reinforcement learning algorithm. Simulation results based on the real-world dataset demonstrate that the proposed DT data processing scheme outperforms benchmark schemes in terms of service latency.
\end{abstract}

\ifCLASSOPTIONonecolumn
	\typeout{The onecolumn mode.}
	\vspace*{-10pt}
\else
	\typeout{The twocolumn mode.}
\fi

\IEEEpeerreviewmaketitle

\MYnewpage


\section{Introduction}
\label{sec:Introduction}


{Short video streaming has significantly altered the landscape of digital content by offering attractable content, aligning with contemporary viewers' preferences for accessible entertainment~\cite{li2023dashlet}. However, seamless video requests put a significant transmission and computing overhead on mobile communication networks~\cite{9374553}.} Multicast transmission, as an efficient data transmission method, allows video data disseminated to multiple users simultaneously over the same wireless channels, which can effectively alleviate the network overhead~\cite{report}. To achieve low-latency \ac{msvs}, efficient network management is necessary, including accurate \ac{mg} update, adaptive video transcoding, and high-throughput multicast transmission.

\Ac{dt} is an effective method to achieve efficient network management through its advanced network status emulation, data feature abstraction, and network decision-making capabilities~\cite{holi,huang2023}. For instance, \ac{dt} can use prediction-based algorithms, such as \ac{lstm} and \ac{rnn}, to emulate the real-time network status of \ac{mg}, which can effectively reduce data collection cost~\cite{granelli2021evaluating}. The emulated network status can be further analyzed by the data abstraction module in \ac{dt} to mine user similarity for accurate \ac{mg} update. Furthermore, the \ac{mg}'s buffer dynamics can be mimicked by \ac{dt} through the embedded swipe-aware buffer update methods, which can facilitate adaptive video transcoding to avoid the buffer from being empty. Additionally, the data-model-driven network management algorithms in \ac{dt} can handle the complex segment buffering and communication resource allocation problems to achieve high-throughput multicast transmission. {However, since the complex embedded algorithms for data processing in \ac{dt} can bring huge computation cost to the mobile communication networks, it is necessary to optimize \ac{dt} data processing configurations to improve its efficiency.}

{However, achieving efficient \ac{dt} data processing for low-latency \ac{msvs} faces several challenges. Firstly, since \ac{dt} data processing methods usually include machine learning methods that are resource-intensive, always using the static neural network structure and size to process the dynamic network status can cause the inefficiency of model inference.} Therefore, how to select an appropriate \ac{dt} model size to adapt to network dynamics is challenging. Secondly, considering the \ac{dt} data processing accuracy of user clustering and abstracted swipe features can affect the transcoding and transmission efficiency, how to characterize the impact is challenging. Thirdly, since the network management problem for \ac{msvs} is usually a mixed-integer nonlinear programming problem, how to design an efficient algorithm to solve it is challenging.

We propose a novel efficient \ac{dt} data processing scheme, which can efficiently reduce service latency for \ac{msvs}. The main contributions are summarized as follows:
\begin{itemize}
    \item {Firstly, we propose a precise measurement model of \ac{dt} data processing to characterize the relationship among DT model size, user dynamics, and user clustering accuracy.} Specifically, the model size is analyzed based on the weights and biases of each layer in the neural networks as well as the number of centroids. The data variation of user status is utilized to represent the user dynamics. A quadratic polynomial is used to fit the mathematical relationship with a low error.
    \item Secondly, we propose a novel service latency model by incorporating the impact of user clustering accuracy. Specifically, the user clustering accuracy affects users’ average engagement time, which further causes different transmission resource demands. Furthermore, considering the sequential process between video transcoding and multicast transmission, a refined service latency model is established.
    \item Thirdly, we propose a diffusion-based resource management algorithm. The denoising process of the diffusion model can improve the action-generation process of \ac{drl} algorithm even if the environment is complex and the action dimension is high. The extensive simulation results on real-world short video streaming datasets show that the proposed diffusion-based resource management algorithm can effectively reduce service latency compared with benchmark schemes.
\end{itemize}

The remainder of this paper is organized as follows. The system model is first built in Section II, followed by the efficient \ac{dt} data processing scheme in Section III. Then, the problem formulation and the proposed scheduling algorithm are shown in Section IV, followed by simulation results in Section V. Finally, Section VI concludes this paper.

\section{System Model}

\begin{figure}[!t]
    \centering
    \includegraphics[width=0.7\mysinglefigwidth]{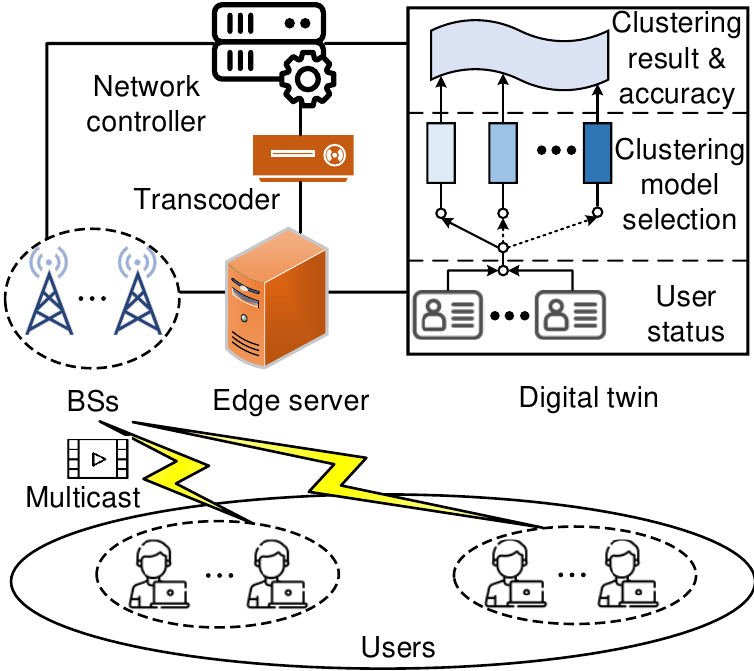}
    \caption{DT-assisted MSVS scenario.}
    \label{syst}
\end{figure}

As shown in Fig.~\ref{syst}, we consider a DT-assisted MSVS scenario, which mainly consists of multiple \acp{bs}, \acp{mg}, one edge server equipped with a transcoder for video transcoding, one \ac{dt}, and one network controller (NC).

\begin{itemize}
    \item Base station: \acp{bs} are responsible for transmitting video sequences to users through multicast transmission, which can effectively reduce the network traffic burden. 
    \item Multicast group: The \ac{mg} consists of multiple users using the short video streaming service. Due to the similarity of users’ preferences and locations, the same short video sequences can be multicast to one \ac{mg}. The user set of \ac{mg} $g$ is denoted by $\mathcal{K}_g$, and the number of \acp{mg} is $G$. The recommended video list of MG $g$ is $\Delta_g$.
    
    \item Edge server: {The edge server owns limited caching and computing capability, which usually stores popular short video sequences in fast response to users’ requests. The edge server is also equipped with a transcoder for real-time video transcoding, which can dynamically adjust the video bitrate to reduce playback lags.}
    \item Digital twin: \ac{dt} is deployed at the edge server, which stores users’ status data, such as locations, channel conditions, swipe timestamps, and preferences. The data is mainly generated through the \ac{lstm} model in \ac{dt} by training the collected historical user status. Based on the stored data in \ac{dt}, the improved user clustering model in \ac{dt} can analyze user similarity for accurate \ac{mg} update~\cite{jstsp}, but consumes plenty of computing resources even if the user status is relatively static. Therefore, we deploy multiple user clustering models of various sizes in \ac{dt} and adaptively select an appropriate one based on user dynamics to reduce DT data processing delay.

    \item Network controller: NC decides the bandwidth allocation and DT model selection based on scheduling algorithms.
\end{itemize}

{The \ac{dt}-assisted MSVS operates as follows. \ac{dt} first generates future user status in the next large resource reservation window and analyzes user dynamics. Then, an appropriate user clustering model is activated to implement \ac{mg} update and output the clustering accuracy. Next, the swipe probability distribution and recommended video list are abstracted from each \ac{mg} to analyze the service latency by integrating allocated bandwidth and computing resources, where the clustering accuracy affects the efficiency of multicast transmission. Finally, the short video sequences are transcoded to appropriate bitrates and transmitted to \acp{mg}.}

\section{Efficient DT Data Processing Scheme}
In this section, we first construct \ac{dt}, and then model the impact of \ac{dt} data processing on service latency.
\subsection{Digital Twin Construction}

\subsubsection{DT data acquisition}
DT data mainly consists of realistic user status data collected from the physical mobile communication network and emulated user status data from the embedded prediction-based algorithms. Specifically, DT data can be classified into networking-related data and behavior-related data. For example, the networking-related data includes users' channel conditions, base stations' transmission capabilities and traffic load, edge servers' caching and computing workload, etc. The behavior-related data includes users' swipe frequency, locations, preferences, etc. Based on the collected data via \acp{bs}, DT can conduct status emulation based on the embedded prediction-based algorithms, such as \ac{lstm} and \ac{rnn}, which can effectively reduce data collection overhead in the physical mobile communication network. 

\subsubsection{DT data processing}
\begin{figure*}[!t]
    \centering
    \includegraphics[width=1.7\mysinglefigwidth]{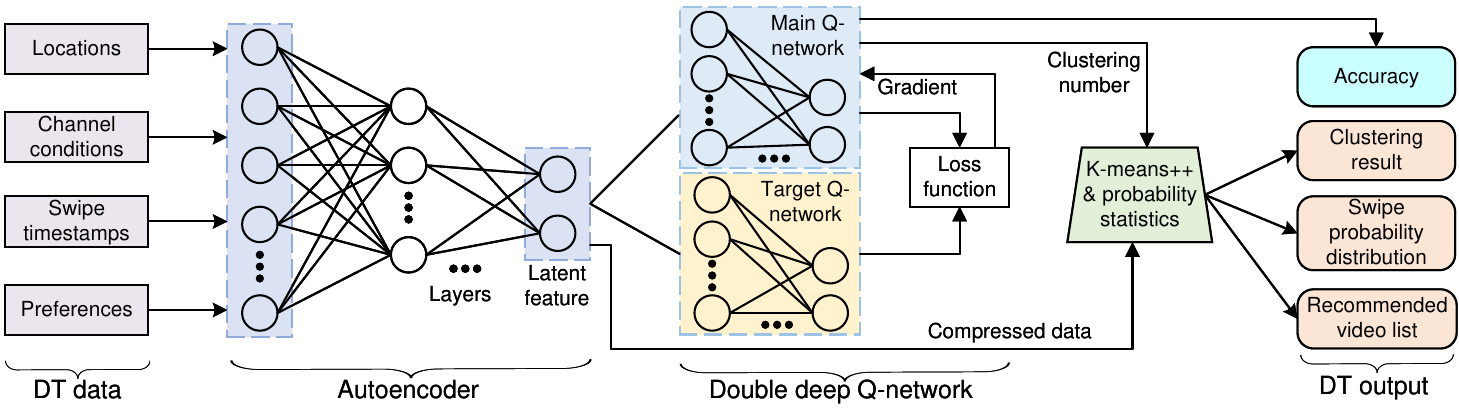}
    \caption{DT data processing procedure.}
    \label{pro}
\end{figure*}

As shown in Fig.~\ref{pro}, we present a detailed DT data processing procedure. The DT data processing module mainly consists of three parts, i.e., autoencoder, \ac{ddqn}, and K-means++ with probability statistics. Specifically, the autoencoder is responsible for compressing time-series DT data into low-dimensional latent features. The \ac{ddqn} analyzes the latent features to determine the appropriate clustering number and output model accuracy~\cite{ddqn}. {Here, the accuracy refers to the probability of the action output. Based on the compressed DT data and clustering number, the K-means++ method can realize fast user clustering while the probability statistics can output the swipe probability distribution and recommended video list for each \ac{mg}~\cite{jstsp}.}

In this procedure, the autoencoder and \ac{ddqn} usually occupy the majority of computing workload due to the hidden neural network layers. Considering the fluctuation of user dynamics, the shallow autoencoder and \ac{ddqn} can be used to process \ac{dt} data when the user status is relatively static, which can effectively reduce DT data processing consumption.

{Based on the abovementioned analysis, we further model the mathematical relationship among DT data processing model size selection, user dynamics, and clustering accuracy.} We first analyze the user dynamics in resource reservation window $t$. The user status in DT consists of four aspects, i.e., locations, $\mathbf{Y}_{t}$, channel conditions, $\mathbf{H}_{t}$, swipe timestamps, $\mathbf{W}_{t}$, and preferences, $\mathbf{E}_{t}$. The above user status components are vectors consisting of multiple time-series data in large resource reservation window $t$. We utilize the data variation to represent the overall user dynamics, ${{\psi }_{t}}$, which can be expressed as:
\begin{align}
    {{\psi }_{t}}={\delta }_{1}\operatorname{var}\left( {\mathbf{H}_{t}} \right)+{{\delta }_{2}}\operatorname{var}\left( {\mathbf{Y}_{t}} \right)+{{\delta }_{3}}\operatorname{var}\left( {\mathbf{W}_{t}} \right)+{{\delta }_{4}}\operatorname{var}\left( {\mathbf{E}_{t}} \right),
\end{align}
where ${{\delta }_{1}},{{\delta }_{2}},{{\delta }_{3},{\delta }_{4}}$ represent the weighting factors of different user status components for normalization.

To guarantee the flexibility and effectiveness of \ac{dt} data processing model, we set three kinds of model versions by adjusting the number of hidden neural network layers. The model selection variable is denoted by $a_i$, where model version $i$ ranges from $1$ to $3$. Here, $a_i$ is a binary variable, representing if model version $i$ is activated, $a_i=1$; otherwise, $a_i=0$. The model size of model version $i$ is denoted by $m_i$, which mainly includes the weights and biases of each layer in the autoencoder and \ac{ddqn} as well as the number of centroids, $K$, and data dimensionality, $d$, in the K-means++, as follows: 
\begin{align}
{{m}_{i}}&={{S}_{\text{p}}}\left( \sum\nolimits_{\texttt{Conv2D}}{\left( w_{i}^{\text{C}}+b_{i}^{\text{C}} \right)}+\sum\nolimits_{\texttt{Dense}}{\left( w_{i}^{\text{D}}+b_{i}^{\text{D}} \right)} \right)\notag\\&+K\times d\times {{S}_{\text{d}}},
\end{align}
where $w_i^{\text{C}},w_i^{\text{D}}$ are the number of weights in a \texttt{Conv2D} and a \texttt{Dense} layer, while $ b_i^{\text{C}},b_i^{\text{D}}$ are the number of biases in a \texttt{Conv2D} and a \texttt{Dense} layer, respectively. Here, $S_{\text{p}},S_{\text{d}}$ represent the size of each parameter and data feature in megabytes (MB), respectively.
\begin{figure}[!t]
    \centering
    \includegraphics[width=0.75\mysinglefigwidth]{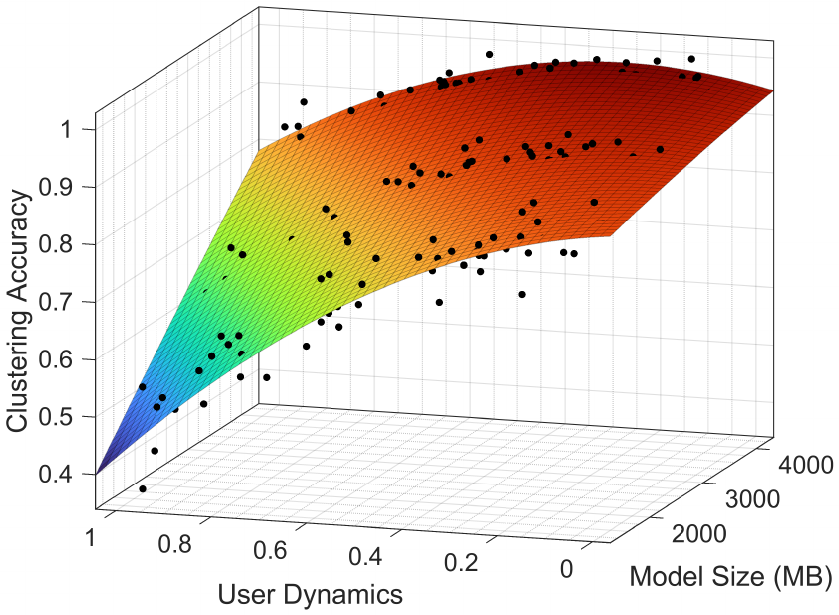}
    \caption{Data fitting on DT data processing accuracy.}
    \label{fitting}
\end{figure}

{We first set three kinds of neural network layers in the DT model and analyze their model sizes. Then, we calculate the user dynamics in different resource reservation windows and measure the corresponding clustering accuracy under different DT model sizes, as shown in Fig.~\ref{fitting}.} It can be observed that with the increasing model size and decreasing user dynamics, the clustering accuracy gradually rises and approaches $1$. The quadratic polynomial in the Curve Fitting Toolbox~\footnote{https://www.mathworks.com/help/curvefit/} at Matlab is utilized to fit the mathematical relationship among DT model size selection, user dynamics, and clustering accuracy, as follows:
\begin{align}
f\left( {{m}_{i}},{{\psi }_{t}} \right)&=0.8246+3.793\times {{10}^{-5}}{{m}_{i}}-0.2262{{\psi }_{t}}\notag\\&-2.044\times {{10}^{-9}}m_{i}^{2}+3.931\times {{10}^{-5}}{{m}_{i}}{{\psi }_{t}}\notag\\&-0.3294\psi _{t}^{2},
\end{align}
where $f\left( {{m}_{i}},{{\psi }_{t}} \right)$ is the clustering accuracy with the model size $m_i$ and user dynamics $\psi_t$. In the above fitting function, the coefficient of determination and \ac{rmse} are $0.9334$ and $0.0326$, which indicate that the fitting function has a high fitting accuracy.

\subsection{Service Latency}
The service latency consists of three parts, i.e., \ac{dt} data processing delay, video transcoding delay, and multicast transmission delay. For the simplification of expression, we omit resource reservation window $t$ in the following modeling.
\subsubsection{\ac{dt} data processing delay}
Since video transcoding and multicast can only be implemented after \ac{dt} data processing, the total computing resources of the edge server can be utilized for \ac{dt} data processing. The corresponding \ac{dt} data processing delay, $\Xi(a_i)$, is expressed as
\begin{equation}
    \Xi\left( {{a}_{i}} \right)=\frac{\kappa {{a}_{i}}{{m}_{i}}}{\mathcal{C}},
\end{equation}
where $\kappa$ is the computing density whose unit is $cycles/MB$. Here, $\mathcal{C}$ is the computing capacity of the edge server.

\subsubsection{Transcoding delay}
To estimate the average transcoding delay in the next resource reservation window, we first need to estimate the average transcoding workload, $\Upsilon_g$, of \ac{mg}~$g$ based on the swipe probability distribution, $p_{g,v}(x)$, and recommended video list in the rest resource window, $\tilde{\Delta}_g(a_i)$, abstracted from \ac{dt}-assisted user clustering, as follows:
\begin{equation}
    {{\Upsilon }_{g}(a_i)}=\mu \sum_{v\in {\tilde{\Delta }_{g}(a_i)}}{\int_{0}^{\varphi \left( v \right)}{\left( 1-{{p}_{g,v}}\left( x \right) \right)\sum\nolimits_{l=2}^{\overline{\mathop{l}}_g\,}{z_{g,v}^{l}\left( x \right)dx}}},
\end{equation}
where $\mu$ and $\varphi(v)$ represent the transcoding density and video length of video $v$. Here, $\tilde{\Delta}_g(a_i)$ is the subset of $\Delta_g$ with the ratio of $\lceil(V-\Xi(a_i))/V\rceil$, where $V$ is the resource reservation window size. {The video sequence is encoded into multiple segments with different layers, $l$, through \ac{svc} technique, indicating that the segment with more layers usually consumes more computing resources but with higher video quality~\cite{9612053}.} Here, $z_{g,v}^{l}(x)$ indicates the layer size. Since the multicast short video version usually needs to adaptively change with users’ dynamic status, we select the average video version~$\overline{l}_g$ of \ac{mg} $g$ in the previous resource reservation window for approximation~\cite{10279382}. 

Since DT data processing occurs before video transcoding, video transcoding can occupy total computing resources in the rest resource reservation window. Here, we consider each MG is allocated to equal computing resources, i.e., $\overline{\mathcal{C}}$. Therefore, the average transcoding delay, ${\Psi}_{g}\left( {{a}_{i}} \right)$, is described as
\begin{equation}
   {{\Psi}_{g}}\left( {{a}_{i}} \right)=\frac{{{\Upsilon }_{g}(a_i)}}{\overline{\mathcal{C}}\sum\nolimits_{v\in\tilde{\Delta}_g(a_i)}\int_{0}^{\varphi(v)}\left(1-p_{g,v}(x)\right)dx}, 
\end{equation}
where the summation integral term in the denominator is the number of estimated segments in the rest resource reservation window.

\subsubsection{Multicast transmission delay}
{Compared with unicast transmission, the multicast transmission capability depends on the user with the worst channel conditions. The accurate user clustering can improve transmission efficiency based on appropriate channel coding rate~\cite{9122408}.} Therefore, the clustering accuracy directly affects the transmission capability, $r_g(a_i,B_g)$, which is given by
\begin{equation}
{{r}_{g}}\left( {{a}_{i}},{{B}_{g}} \right)=f\left( {{a}_{i}}m_i,\psi  \right){{B}_{g}}{{\log }_{2}}\left( 1+\underset{k\in {{\mathcal{K}}_{g}}}{\mathop{\min }}\,\frac{{{\left| {{h}_{g,k}} \right|}^{2}}{{P}_{\text{D}}}}{{{N}_{0}}} \right),
\end{equation}
where $h_{g,k}$ is the channel gain of user $k$ in \ac{mg} $g$. Here, $P_{\text{D}}$ and $B_g$ represent the downlink transmission power and reserved bandwidth resources for \ac{mg} $g$. Additionally, $N_0$ is the noise power.

Based on the transmission capability and transmitted video file size, we can estimate the average transmission delay, $\Gamma_g(a_i,B_g)$, as follows:
\begin{equation}
    {{\Gamma }_{g}}\left( {{a}_{i}},{{B}_{g}} \right)=\frac{\sum\nolimits_{v\in {\tilde{\Delta }_{g}}}{\int_{0}^{\varphi \left( v \right)}{\left( 1-{{p}_{g,v}}\left( x \right) \right)\sum\nolimits_{l=1}^{\overset{-}{\mathop{l}_g}\,}{z_{g,v}^{l}\left( x \right)dx}}}}{{{r}_{g}}\left( {{a}_{i}},{{B}_{g}} \right)\sum\nolimits_{v\in\tilde{\Delta}_g(a_i)}\int_{0}^{\varphi(v)}\left(1-p_{g,v}(x)\right)dx},
\end{equation}
where the numerator is the estimated transmitted video file size by integrating swipe probability distribution, recommended video list, and average video version.

Based on the above analysis, we can estimate the service latency, $\mathcal{T}$, by integrating \ac{dt} data processing delay, transcoding delay, and multicast transmission delay, as follows:
\begin{equation}
    \mathcal{T}\left(\mathbf{a},\mathbf{B} \right)=\sum\limits_{i=1}^{I}{\Xi \left( {{a}_{i}} \right)+\frac{1}{G}\sum\limits_{i=1}^{I}{\sum\limits_{g=1}^{G}\left({{{\Psi}_{g}}\left( {{a}_{i}} \right)+{{\Gamma }_{g}}\left( {{a}_{i}},{{B}_{g}} \right)}\right)}},
\end{equation}
where $\mathbf{a}=\{a_i\}_{i\in\mathcal{I}}$ and $\mathbf{B}=\{B_g\}_{g\in\mathcal{G}}$ are the model selection variable set and reserved bandwidth variable set, respectively.

\section{Problem Formulation and Solution}
\subsection{Problem Formulation}
Since network resources are limited, efficient DT data processing and multicast transmission strategies can guarantee \ac{msvs} with a low service latency. In each resource reservation window, our objective is to minimize the service latency by adjusting DT model selection and bandwidth reservation. The corresponding optimization problem is formulated as
\begin{align}\label{opt}
 \textbf{P}_0: & \underset{\left\{ {\mathbf{a}_{t}},{\mathbf{B}_{t}} \right\}}{\mathop{\min }}\,\underset{T\to \infty }{\mathop{\lim }}\,\frac{1}{T}\sum\nolimits_{t=1}^{T}{{{\mathcal{T}}_{t}}\left( {\mathbf{a}_{t}},{\mathbf{B}_{t}} \right)} \\ 
  \text{s.t.}&\sum\nolimits_{i=1}^{I}{{{a}_{t,i}}}=1,{{a}_{t,i}}\in \left\{ 0,1 \right\},\tag{10a} \\ 
 & \sum\nolimits_{g=1}^{G}{{{B}_{t,g}}}\le \mathcal{B},{{B}_{t,g}}\in \left[ 0,\mathcal{B} \right].\tag{10b}
\end{align}
Constraint (10a) guarantees the efficiency of DT data processing by only selecting one DT model for DT data processing in resource reservation window $t$. Constraint (10b) guarantees the bandwidth resources reserved for all \acp{mg} do not exceed the total bandwidth capacity, $\mathcal{B}$.

\subsection{Diffusion-Based Resource Management Algorithm}
Since the formulated problem is mixed-integer nonlinear programming, which is hard to solve directly. Considering that the \ac{mg}’s playback status satisfies Markov chain and the optimization objective is to minimize the long-term service latency, the optimization problem can be modeled as a \ac{mdp}. {The commonly used algorithm is \ac{td3} to solve the \ac{mdp}~\cite{td3paper}, but it hardly converges to a high reward with less fluctuation when the environment is complex and the action dimension is high.} Therefore, we propose a \ac{dt3} algorithm to improve the action-generation process of \ac{td3} through the denoising process of the diffusion model~\cite{ajay2022conditional}.

\subsubsection{State}
The state consists of \acp{mg}’ transcoding workload, DT data processing delay, and average transcoding and multicast transmission delay, as follows:
\begin{equation}
    {{S}_{t}}=\left\{ {{\left\{ {{\Upsilon }_{t,g}} \right\}}_{g\in \mathcal{G}}}, \Xi_t, \left\{\Psi_{t,g} \right\}_{g\in \mathcal{G}}, \left\{\Gamma_{t,g} \right\}_{g\in \mathcal{G}}\right\}.
\end{equation}
\subsubsection{Action}
The action includes all optimization variables in Eq.~\eqref{opt}, which is defined by
\begin{equation}
    {{A}_{t}}=\left\{ {\mathbf{a}_{t}},{\mathbf{B}_{t}} \right\}.
\end{equation}
\subsubsection{Reward}
The reward at step $t$ is designed to maximize the opposite of the overall service latency, which is defined by
\begin{equation}
    {{R}_{t}}=-\mathcal{T}\left( {\mathbf{a}_{t}},{\mathbf{B}_{t}} \right).
\end{equation}

\subsubsection{Diffusion-based action generation}
In each time step $t$, the model selection and bandwidth resource reservation decisions are conducted based on the current state $S_t$, which can be described as $\pi_{\theta}(A_t|S_t)$. The diffusion process is to add a sequence of Gaussian noises to a target probability distribution, $x_0$, at each denosing step $j$, which can be expressed as
\begin{equation}\label{rev}
    q\left( {{x}_{j}}|{{x}_{j-1}} \right)=\mathcal{N}\left( {{x}_{j}};\sqrt{1-{{\beta }_{j}}}{{x}_{j-1}},{{\beta }_{j}}\mathbf{I} \right),
\end{equation}
where $\beta_j$ and $\sqrt{1-{{\beta }_{j}}}{x}_{j-1}$ represent the forward process variance and the mean of the normal distribution, respectively.

Based on the diffusion process, we can utilize the reverse process of conditional diffusion models to model the actor’s policy~\cite{liu2024ddm}, as follows:
\begin{align}\label{act}
    {{A}^{j-1}}|{{A}^{j}}=\frac{{{A}^{j}}}{\sqrt{{{\alpha }^{j}}}}-\frac{{{\beta }_{j}}}{\sqrt{{{\alpha }^{j}}\left( 1-{{\overline{\alpha }}^{j}} \right)}}{{\varepsilon }_{\theta }}\left( {{A}^{j}},S,j \right)+\sqrt{{{\beta }_{j}}}\varepsilon,
\end{align}
where $\alpha^j$ equals to $1-\beta_j$ and $\varepsilon \sim \mathcal{N}\left( \mathbf{0},\mathbf{I} \right)$ is a standard normal noise. Here, the denoising step is denoted by $j$, ranging from $1$ to $N$.

\subsubsection{Actor-critic network training}
The training process involves several key techniques to effectively learn optimal or near-optimal policies in complex environments. {First, \ac{td3} utilizes two separate neural networks, i.e., the actor and critic networks, where the former is responsible for selecting actions, while the latter evaluates the quality of those actions.} Then, \ac{td3} employs a form of off-policy learning, where it samples experiences from a replay buffer. Finally, one distinguishing feature of \ac{td3} is the use of twin critics. These twin critics help mitigate overestimation bias in the Q-value estimates, which can generate more stable and accurate evaluations of actions. The actor network is updated by the deterministic policy gradient~\cite{td3paper}, as follows:
\begin{equation}\label{update}
    {{\nabla }_{\phi }}J\left( \phi  \right)=\frac{1}{\Im }\sum{{{\nabla }_{A}}{{Q}_{{{\theta }_{1}}}}\left( S,A \right){{\nabla }_{\phi }}{{Q}_{\phi }}\left( S \right)}
\end{equation}
where $\Im $ is the number of mini-batch size. Here, $\theta_1$ and $\phi$ represent the network parameters of critic and actor networks, respectively. {Since the generated action by \ac{td3} is the continuous value, we split the action range of model selection variable $\mathbf{a}$ into $I$ parts, where the largest part is selected as the \ac{dt} model version.} The detailed algorithm procedure of proposed \ac{dt3} is presented in Algorithm~\ref{alg1}.

\begin{algorithm}[t]\label{alg1}
\caption{Diffusion-Based TD3 (DFTD3). }
\textbf{Input:} Computing capacity $\mathcal{C}$, bandwidth capacity $\mathcal{B}$, DT model relationship $f(m_i,\psi)$, and users’ channel gains $\mathbf{H}$;

\textbf{Output:} Model selection variable $\mathbf{a}$ and reserved bandwidth variable $\mathbf{B}$;

\textbf{Initialize:} Critic networks $Q_{\theta_1}$, $Q_{\theta_2}$ and actor network $\pi_{\phi}$ with random parameters $\theta_1$, $\theta_2$, $\phi$; Target networks $\theta_1^{'} \leftarrow \theta_1$, $\theta_2^{'} \leftarrow \theta_2$, $\phi^{'} \leftarrow \phi$; Relay buffer $\mathcal{R}$;

\For{each $\text{episode}$}
    {
    Reset initial state $s_1$;
    
        \For{\text{each step} $t \in \{1,\cdots,T\}$}
            {
            Observe state $S$ and initialize a random normal distribution $x_t\sim\mathcal{N}(\mathbf{0},\mathbf{I})$;
            
            \For{denoising step $j=t : 1$}{
                Calculate the reverse transition distribution based on Eq.~\eqref{rev};

                Generate action $A$ based on Eq.~\eqref{act};                
            }

            Execute action ${A}$ and observe reward $R$;

            Store transition tuple $(S,A,R,S')$ in $\mathcal{R}$;

            Update critic network parameters: $\theta_n\leftarrow \min_{\theta_n}\Im^{-1}\sum(y-Q_{\theta_n}(S,A)^2$;

            \If{$t$ mod $10$}{
            Update actor network parameter: $\phi$ based on Eq.~\eqref{update};

            Update target network parameters: $\theta_n^{'}\leftarrow \tau\theta_n+(1-\tau)\theta_n^{'}$, $\phi^{'}\leftarrow \tau\phi+(1-\tau)\phi^{'}$;
            }          
            }
    }
\end{algorithm}

\section{Simulation Results}

\subsection{Simulation Setup}
We conduct extensive simulations on the real-world video streaming dataset to evaluate the performance of the proposed \ac{dt3} algorithm. The main simulation parameters are presented in Table~\ref{table_1}. The key simulation components of how to construct \ac{dt} are introduced as follows.

\begin{table}[!t]
\centering
\caption{Main Simulation Parameters}
\label{table_1}
\begin{tblr}{
    width=\linewidth,
    colspec = {X[1,c,m]X[1.8,c,m]X[1,c,m]X[2.2,c,m]},
    hlines,
    hline{2} = {1}{-}{},
    hline{2} = {2}{-}{},
    vline{2},
    vline{3} = {1}{-}{},
    vline{3} = {2}{-}{},
    vline{4},
    column{2,4} = {mode=dmath},
}
\textbf{Parameter} & \textbf{Value} &\textbf{Parameter} & \textbf{Value} \\
$\mathcal{C}$ & [6,10]~\myunit{Gcycles/s} & $V$ & 5~\myunit{min} \\
$\mathcal{B}$ & [6,14]~\myunit{MHz} & $P_{\text{D}}$ & 27~\myunit{dBm} \\
$\kappa$ & 16~\myunit{Mcycles/Mb} & $N_0$ & -174~\myunit{dBm}\\
$\mu$ & 6~\myunit{Gcycles/Mb} & $\bm{\delta}$ & \{0.125,0.25,0.8,0.1\}
\end{tblr}
\end{table}

\begin{figure*}[!t]
    \centering
    \subfloat[Convergence performance of proposed algorithm.]{\includegraphics[width=0.33\linewidth]{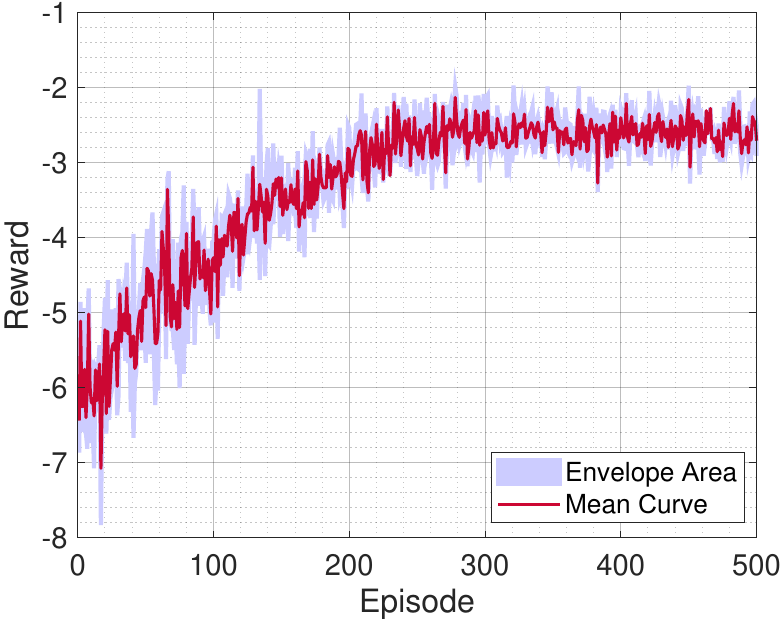}\label{fig_sub1}}
    \hfill 
    \subfloat[Service latency vs. bandwidth.]{\includegraphics[width=0.33\linewidth]{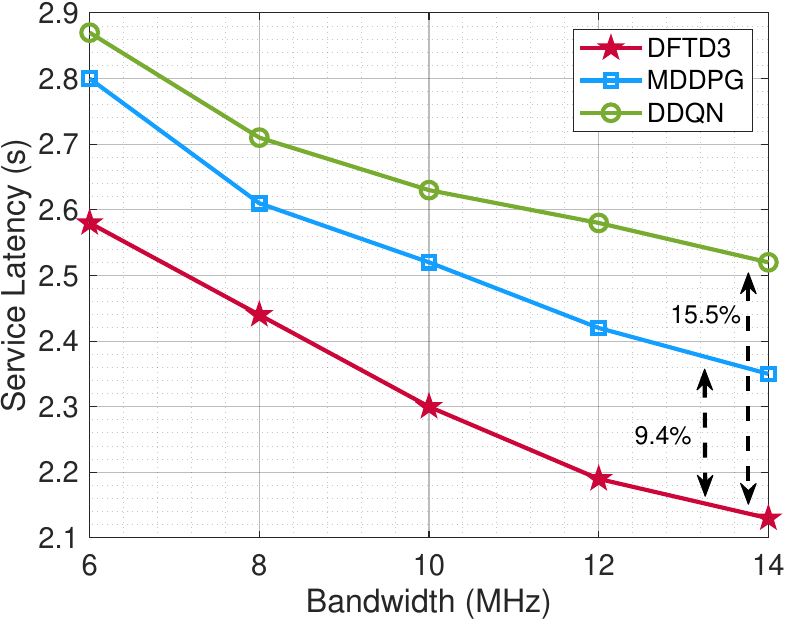}\label{fig_sub2}}
    \hfill 
    \subfloat[Service latency vs. computing capacity.]{\includegraphics[width=0.33\linewidth]{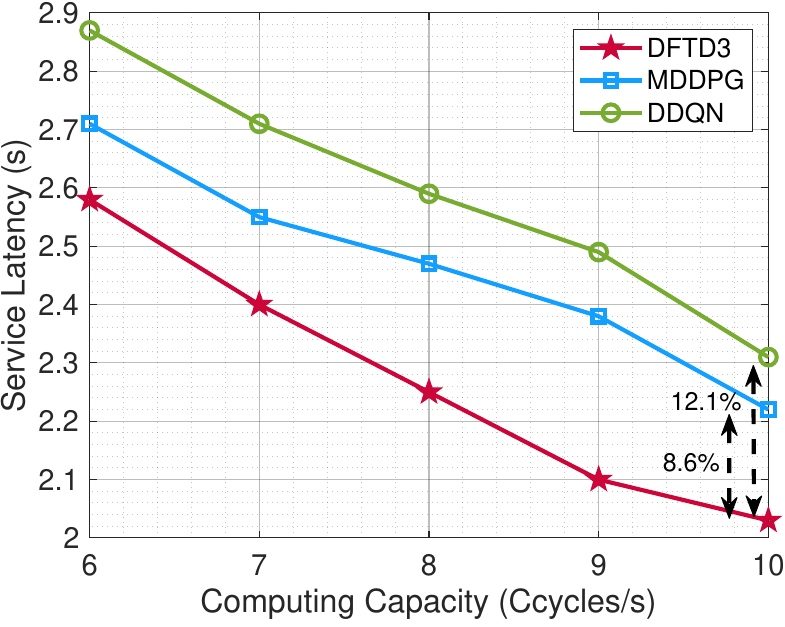}\label{fig_sub3}}
    \caption{Performance evaluation on convergence performance and service latency  .}
    \label{performance}
\end{figure*}

To construct \ac{dt}, we first generate sixty users' trajectories within the University of Waterloo (UW) campus with the Levy flight model that refers to a random Markovian walk and the probability distribution of step lengths satisfies heavy-tailed distribution. The generated data is used as the label data, where eighty percent of label data is selected as the training sample to train the LSTM model for users' trajectory emulation. Based on users' real-time trajectories, the real-time channel conditions are emulated based on PropagationModel at Matlab. Additionally, We adopt the short video streaming dataset\footnote{ACM MM Grand Challenges: https://github.com/AItransCompetition/Short Video-Streaming-Challenge/tree/main/data} to generate users’ swipe behaviors, and sample 1000 short videos from the YouTube 8M dataset\footnote{YouTube 8M dataset: https://research.google.com/youtube8m/index.html}, including 8 video types, i.e., Entertainment, Games, Food, Sports, Science, Dance, Travel, and News. Each video has a duration of 15 $sec$ and is encoded into four versions. 

We compare the proposed \ac{dt3} algorithm with the following benchmark algorithms:
\begin{itemize}
    \item \textbf{Modified deep deterministic policy gradient (MDDPG)-based resource management~\cite{10001020}:} The state, action, and reward function are the same as that in \ac{dt3} algorithm. However, the output value of the Tanh function from $[-1,1]$ is mapped to $[0,1]$. The range of model selection variable $\mathbf{a}$ is split into $I$ parts, where the largest part is selected as the \ac{dt} model version. 
    \item \textbf{Dueling deep Q-network (DDQN)-based resource management~\cite{10279382}:} The action space of bandwidth reservation consists of $10$ sub-actions, where each of them is a binary value of $0$ or $1$. The state and reward function are the same as that in \ac{dt3}.
\end{itemize}

\subsection{Performance Evaluation}
As shown in Fig.~\ref{performance}(a), we conduct two training trails to draw the corresponding envelope curve and the mean curve. The reward of the proposed DFTD3 algorithm can converge to a stable value around $-2.6$ when the episode reaches about $240$. This demonstrates that the diffusion model can effectively improve the action-generation process and the refined TD3 algorithm can explore a good resource management strategy. Furthermore, as shown in Fig.~\ref{performance}(b), it can be observed that the service latency gradually decreases with the increasing bandwidths and computing capacities while the proposed DFTD3 algorithm always maintains the lowest service latency compared with other benchmark schemes, which implies that it can well adapt to varying network capacities.

\section{Conclusion}
\label{sec:Conclusion}

In this paper, we have proposed an efficient DT data processing scheme to reduce service latency for MSVS. A diffusion-based resource management algorithm has been designed to improve network decision-making performance. The proposed DT data processing scheme can well adapt to different user dynamics. For the future work, we will investigate the adaptive DT model update mechanism to further improve DT data processing efficiency.

%
\bibliographystyle{IEEEtran}
\bibliography{IEEEabrv,Ref}
%
%

\end{document}